\documentclass[12pt,draftcls,onecolumn]{IEEEtran}
\usepackage[pagewise,switch]{lineno}
\usepackage{booktabs}
 \usepackage{multirow}
\usepackage{times}
\usepackage[ruled,linesnumbered,vlined]{algorithm2e}
\usepackage{cite}
\usepackage{indentfirst}
\usepackage{graphicx}
\usepackage{subfigure}
\usepackage{amsmath}
\usepackage{multirow}
\usepackage{float}
\usepackage{soul}
\usepackage{color}
\usepackage{subfigure}
\usepackage{graphicx}
\usepackage{caption}
\usepackage{multirow}
\usepackage{amsthm}
\usepackage{url}

\usepackage{ulem}
\usepackage{verbatim}
\usepackage{tcolorbox}
\begin{document}

\title{SleepGuardian: An RF-based Healthcare System Guarding Your Sleep from Afar}

\author{Yu~Gu,~\IEEEmembership{Senior Member,~IEEE,}
        Yantong~Wang,
        Zhi~Liu,~\IEEEmembership{Member,~IEEE,}
        Jun~Liu,~\IEEEmembership{Member,~IEEE,}
        Jie~Li,~\IEEEmembership{Senior Member,~IEEE,}

\IEEEcompsocitemizethanks{
\IEEEcompsocthanksitem Y. Gu, Y. Wang, and J. Liu are with School of Computer and Information, Hefei University of Technology, China.
E-mail: hfut\_bruce@hfut.edu.cn
\IEEEcompsocthanksitem Z. Liu (corresponding author) is with Shizuoka University, Japan
E-mail: liu@ieee.org
\IEEEcompsocthanksitem J. Li is with Dept. of Computer Science, Shanghai Jiaotong University, China.
E-mail: lijie@cs.sjtu.edu.cn
}}
\maketitle
\markboth{IEEE Network Magazine}%
{Gu \MakeLowercase{\textit{et al.}}: \textcolor{Blue} {An RF-based Healthcare System Guarding Your Sleep from Afar}}

\begin{abstract}
The ever accelerating process of urbanization urges more and more population into the swelling cities. While city residents are enjoying an entertaining life supported by advanced informatics techniques like 5G and cloud computing, the same technologies have also gradually deprived their sleep, which is crucial for their wellness. Therefore, sleep monitoring has drawn significant  attention from both research and industry communities. In this article, we first review the sleep monitoring issue and point out three essential properties of an ideal sleep healthcare system, i.e., realtime guarding, fine-grained logging, and cost-effectiveness. Based on the analysis, we present SleepGuardian, a Radio Frequence (RF) based sleep healthcare system leveraging signal processing, edge computing and machine learning. SleepGuardian offers an offline sleep logging service and an online abnormality warning service. The offline service provides a fine-grained sleep log like timing and regularity of bed time, onset of sleep and night time awakenings. The online service keeps guarding the subject for any abnormal behaviors during sleep like intensive body twitches and a sudden seizure attack.  Once an abnormality happens, it will automatically warn the designated contacts like a nearby emergency
room or a closeby relative. We prototype SleepGuardian with low-cost WiFi devices and evaluate it in real scenarios. Experimental results demonstrate that SleepGuardian is very effective.

\end{abstract}

\section{Introduction}
\label{Sect:Introduction}

Sleep is crucial for the wellness of human being. According to the latest report of the U.S. Bureau of Labor Statistics (BLS), employed persons aging from 25 to 54 spend an average of $8.8$ hours working and $7.8$ hours sleeping \cite{BLS2016}. During sleep, human body remains in an anabolic state to restore both the muscular and nervous systems vital to the body and cognitive function. Individuals with sleep time deviating from the population norm are at risks of various sleep disorders like Sleep Disorder Breathing (SDB), Sleep Behaviour Disorder (SBD), Restless Leg Syndrome (RLS), and Periodic Limb Movement in Sleep (PLMS). These disorders may be serious enough to interfere with normal physical, mental, social and emotional functioning, especially hazardous for elderly.


\subsection{Sleep Monitoring}
Sleep disorders are shown to be diagnosable via a fine-grained sleep history log represented in terms of \emph{still postures} and \emph{in-place motions} \cite{Seelye2015The}. The log contains information like timing and regularity of bed time, onset of sleep, night time awakenings, time of waking up in the mornings, day time naps, and day time sleepiness, etc. It is very challenging to acquire such long-term logs since manual collection is quite time-consuming and labor-intensive. Therefore, the automatic sleep monitoring system rises as an emerging topic and draws more and more  attention from both academic and industrial communities.

\begin{figure}
    \centering
  \includegraphics[width=\columnwidth]{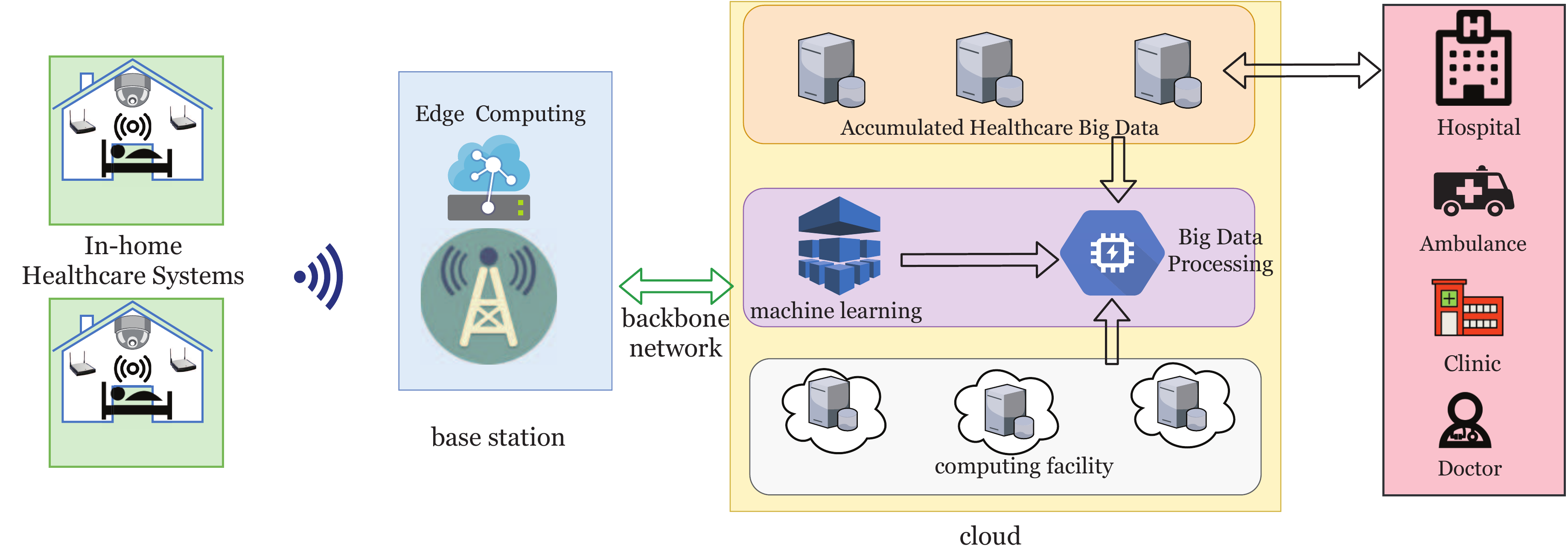}
    \caption{Our proposed framework to bridge between users and social healthcare services. Distributed in-home healthcare systems are connected through base stations which leverage the edge computing to ensure prompt medical response for emergency.  The accumulated healthcare big data in the clouds demanding for big data anaylsis for better diagnosis and treatment. }
   \label{fig1}
\end{figure}

We present a representative framework for an ideal healthcare system leveraging cloud computing, edge computing, and big data processing in Figure \ref{fig1}. An ideal system should bridge between users and social healthcare services in a realtime manner. Healthcare data harvested by distributed and heterogeneous in-home healthcare systems should be timely transmitted, stored and processed in clouds via either wired or high-speed wireless networks like 5G, depending on specific circumstances. Therefore, cloud computing, which relies on the rich computing power on the cloud, emerges as an efficient and reliable approach. Moreover, edge computing is also needed in time-sensitive applications, since edge computing will greatly accelerate processing the data at the proximity and thus ensure prompt medical response for users. Big data analysis on the collected healthcare data is essential for better diagnosis and treatment. In particular, an ideal sleep monitoring solution should possess the following properties,

\textbf{Realtime Sleep Guarding.}  The system should provide a realtime guarding service so that any abnormalities, defined upon applications, can be captured, reported and handled in time. 

\textbf{Fine-grained Status Logging.} A fine-grained sleep history log should be automatically and accurately generated and recorded, so that it can be promptly analyzed to aid diagnosis.

\textbf{Cost-efficiency and Privacy-preserving.} Sleep monitoring is long-term in essence. To maintain a continuous and fine-grained sleep log for accurate diagnosis, it is unreasonable to rely only on the monitoring service provided by hospitals. Therefore, the system should be cost-effective and affordable for general public to support daily usage. Moreover, the privacy issue is another major concern, since current mainstreaming sensor-based solutions are contact or even invasive, which could make users unconformable during sleep and lead to biased data.

\subsection{Current Solutions}
Actigraphy (ACT) has been accepted as a valid tool for this sleep history logging problem by the American Sleep Disorders Association (ASDA) at 1995  \cite{Sadeh1995The}. It refers to methods using wristband-like devices to monitor and collect data generated by body movements. Several such commercial products are already available on the open market such as Mi Band and Fitbit. Actigraphy is considered as a low-cost solution that can provide a fine-grained sleep log. However, due to the volume and capacity constrains, it usually possesses little realtime communication and processing abilities. Moreover, ACT in the shape of wearable devices could be obtrusive for disturbing sleep.

Some recent research try to improve the usability of ACT by attaching sensors to the objects where people lie on rather than the people themselves, e.g., force sensing resistor sensors on the staves of the bed  \cite{Barsocchi2017An} or polyvinylidene fluoride (PVDF) film sensors on a mattress  \cite{Da2014Nocturnal}. But such settings usually have the mobility and coverage issues. Other research like vision systems \cite{Abbas2011Neonatal, Wang2013Unconstrained,Li2016Non} could be contactless but still obstructive due to the coverage issue: illumination and line-of-sight (LOS) restrictions. Also, vision-based approaches inherently raise privacy concerns for the general public.

Radio Frequency (RF) signal that is transparent to users emerges as a new paradigm for sleep monitoring in a ``telepathic'' way \cite{sigg2014telepathic}. Patwari \emph{et al.} presented some pioneering work by exploiting  Received Signal Strength (RSS) of RF signal for extracting respiration rate \cite{Patwari2011Monitoring}. {Since RSS is a coarse-grained indicator that could easily be affected by the varying electromagnetic environment, researchers soon tend to Channel State Info (CSI) of RF signal instead. Liu \emph{et al.} designed Wi-Sleep leveraging multiple pairs of transmitter-receiver antennas for detecting sleep postures and rollovers} \cite{Liu2014WiSleep}. {Later, a similar CSI-based work is proposed to track abnormal breathing like sleep apnea} \cite{Liu2016Contactless}.  Recently we designed Sleepy \cite{Gu2018Sleepy}, a CSI-based realtime healthcare system providing offline status logging.

Existing research during the last few decades has provided solid progresses on this topic. But much effort are still required since most of current solutions possess some major shortcomings facing the above desired properties, e.g, relying on sophisticated hardware, lack of online processing ability, privacy concerns, etc.

\subsection{Our Contributions}
To this end, we extend our previous work Sleepy \cite{Gu2018Sleepy} by endowing it with online processing ability via edge computing and finer recognition granularity via machine learning. The rationale of using edge computing and machine learning is to shorten system response time by keeping computing at the proximity of data, and to enhance system resolution through predictive analytics on data, respectively.  To the best of our knowledge, it is the first-of-its-kind solution fulfilling the aforementioned essential properties for an ideal healthcare system, i.e., realtime guarding, fine-grained logging, cost-effectiveness and privacy-preserving \cite{Hossain2018EnClass}.

The key is to leverage RF signals (more specifically, WiFi) other than ACT or vision as the source for sleep monitoring. RF signal is insensible for human beings, making it an ideal option for this user-centric application. Also, SleepGuardian is cost-effective for using only low-cost off-the-shelf WiFi devices.

But RF signal is usually considered to be vulnerable to the environmental interference caused by either humans or devices. So the main design challenge of SleepGuardian is \textbf{\emph{how to accurately characterize the still posture and in-place motion in terms of RF signal}}?

Our response to the challenge is to build SleepGuardian on a Gaussian Mixture Model (GMM) based approach, which can adaptively model the still posture and in-place motion in Channel State Information (CSI) extracted from WiFi signal sent to monitor a sleeping subject. Specifically, we show that the energy feature of wireless channel follows the GMM when describing the sleep status of a person. Therefore, inspired by the foreground extraction method commonly-used in Computer Vision (CV), we design a similar approach in signal processing where the in-place motion is modeled as foreground and still postures as background in channel response. This approach possesses two major merits: 1) It needs no site-dependent calibrations, and 2) it needs no target-dependent training.

By processing the rich sleep status data offered by GMM, SleepGuardian provides an offline logging service and an online warning service. The offline service provides fine-grained information describing the sleep status of a person in terms of sleeping postures and in-place motions. It can even detect which parts of your body move via machine learning. The online service explores edge computing to guard the subject for any abnormal behaviors like intensive body twitches from a nightmare, a sudden seizure attack or falling off the bed, in a realtime manner. Once an abnormality happens, it will automatically warns the designated contacts like a nearby emergency room or a closeby relative.


We prototype SleepGuardian with low-cost off-the-shelf WiFi devices and verify its performance from different dimensions like the environment, scenario, sampling time, gender, and body shape in terms of overall accuracy (ACC), false negative rate (FNR) and measurement error rate (MER). Experimental results demonstrate that SleepGuardian is very effective and reliable as a low-cost home-use healthcare system for sleep monitoring.

%
%


The remainder of this article is organized as follows. We present our system design of SleepGuardian in the next section, following by the performance evaluation. Finally, we conclude our work.

\section{System Design}
\label{Sect:SysDesign}
\subsection{System Overview}

\begin{figure}
    \centering
     \includegraphics[width=0.6\columnwidth]{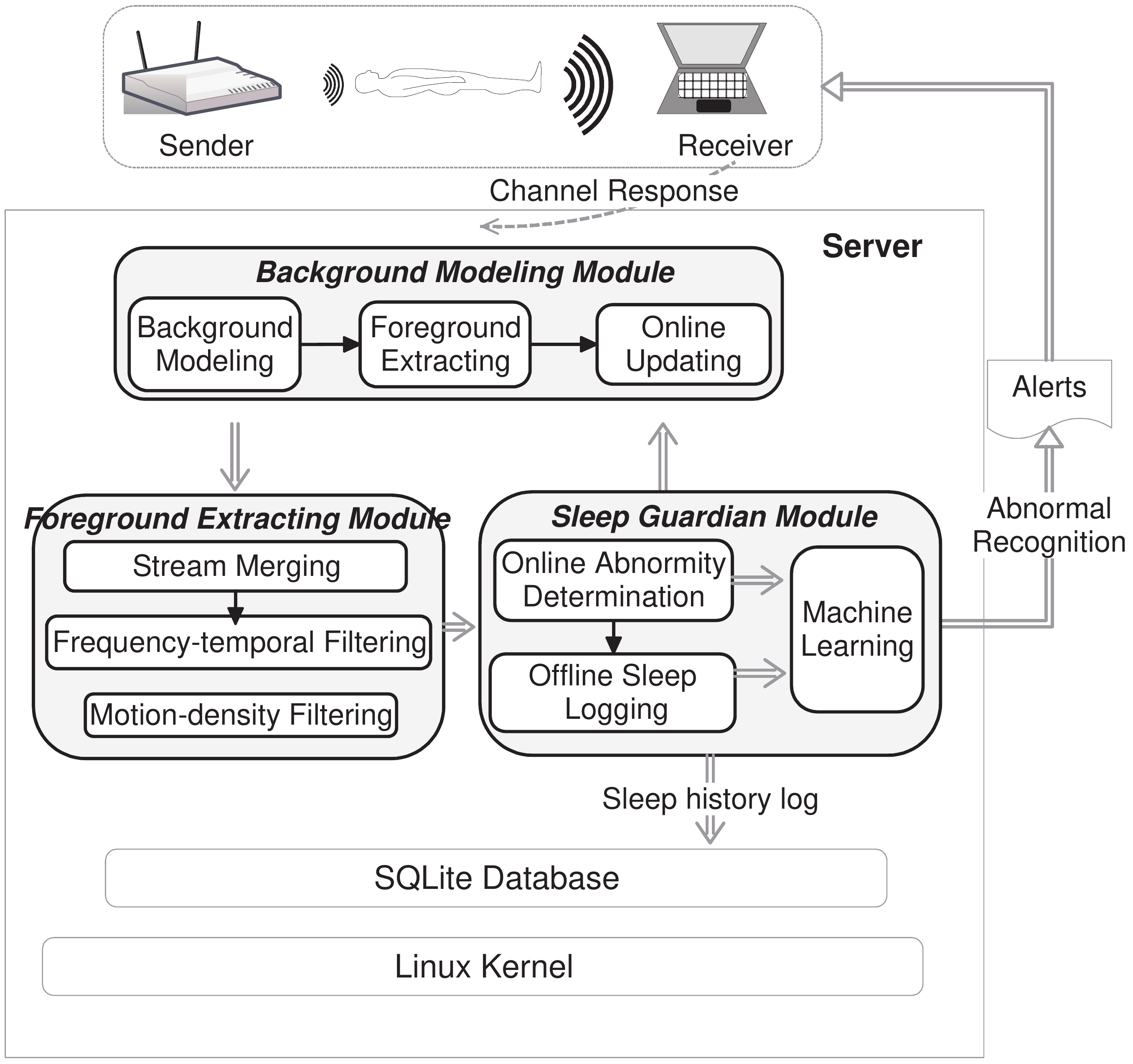}
    \caption{System overview of SleepGuardian.}
   \label{fig:SysArc}
\end{figure}

Figure \ref{fig:SysArc} shows the system architecture of SleepGuardian. The WiFi signal periodically emitted by the sender flows through the target subject and reaches the receiver. The attenuated signal described by channel response contains rich descriptions of the subject like still postures and in-place motions via the multi-path and fading effects. To retrieve fine-grained sleeping information from the signal and offer offline/line monitoring services, the receiver leverages three modules to process CSI data (channel response), i.e., background modeling module, foreground extracting module and sleep guardian module.

\textbf{Background Modeling Module.} This module aims to build an adaptive model for still postures (background) in terms of channel response. It leverages GMM to process raw CSI flows from different receiver antennas.

\textbf{Foreground Extracting Module.} This module targets on extracting in-place motions (foreground) from background for further analysis. It removes counterfeit foreground caused by system glitches and surrounding interferences via the continuity feature of the physical motion in both frequency  and time domain.

\textbf{Sleep Guardian Module.} This module provides the offline logging and online warning services by further processing the background and foreground information provided by the above two modules. More specifically, it logs the duration of different still postures, and the start-time, duration for  the offline services and identifies the abnormal behavior based on the intensity of the in-place motion.

\subsection{Background Modeling Module}
\textbf{ CSI Visualizing}. The incoming CSI data stream should be visualized first. We partition the stream into consecutive windows with length $T$, containing $N$ samples for each subcarrier \cite{Zheng2016Smokey}. Then each frame contains $M\times N$ pixels, where $M$ is the number of subcarriers. {Each pixel $P_{m,n}$ in a frame is the CSI amplitude of subcarrier $m$ collected within the $n$\emph{th} time window ($t_n$) and the color of each pixel is determined by mapping its amplitude value to a predefined colormap.}


\textbf{ Background Modeling}. Within one frame, each pixel is modeled by a mixture of $K$ Gaussian distributions.  The probability that a certain pixel has a value (dBm)  $x_t$ at time $t$ can be written as,

\begin{equation}
p(x_t)=\sum_{i=1}^{K} w_{i,t}\frac{1}{(2\pi)^\frac{n}{2}{|\Sigma|^\frac{1}{2}}} e^{-\frac{1}{2}(x-\mu_{i,it})^T\Sigma^{-1}(x_t-\mu)}
\end{equation}

where $w_{i,t}$ and $\mu_{i,t}$  are the weight and the mean value of the $i$-th Gaussian component at $t$, respectively.

If an incoming pixel does not fit in the background model, it is likely to be the foreground. More specially, if this pixel is more than 2.5 standard deviations away from any of the $B$ distributions in the background model, it is marked as the foreground. 

\textbf{ Online Updating}. Different sleeping postures constitute different backgrounds in our GMM based foreground extraction method. To accurately distinguish foregrounds from changing backgrounds, online updating is essential:

The initial weight and variance of $K$ distributions are set to $\frac{1}{K}$ and 1.5, respectively.

If none of the $K$ distributions match the current pixel, the model is updated by replacing the distribution with \textbf{the least fitness value} with a new Gaussian distribution whose mean is set to the value of current pixel while using an initial weight ($\frac{1}{K}$) and variance (1.5). Then, we normalize the weights of all $K$ distributions.

Otherwise, we update the weight of all $K$ distributions as follows,
\begin{equation}
\begin{split}
w_{i,t}=&(1-\alpha)w_{i,t-1}+ \alpha M_{i,t}, \\
M_{i,t} = &
\begin{cases}
1& \text{if $i$ is the first match distribution}; \\
0& \text{otherwise}
\end{cases}
\end{split}
\end{equation}

\noindent where $\alpha$ is a pre-defined learning rate.

Then, we normalize the weight of all $K$ distributions and update the first matched distribution as follows,
\begin{equation}
\centering
\begin{split}
\mu_{i,t}=&(1-\rho)\mu_{i,t-1}+\rho x_t, \\
\sigma^2_{i,t} = &(1-\rho)\sigma^2_{i,t-1} + \rho(x_t-\mu_{i,t})^2
\end{split}
\end{equation}

where $\rho = \frac{\alpha}{w_{i,t}}$ is a temporary parameter and $\alpha$ is set to 0.01.

The online updating endues our system with the ability to adapt to background changing without extra calibrations or target-dependent training. Detailed information about the GMM model can be found in our previous work \cite{Gu2018Sleepy}.
%
%
%


\subsection{Foreground Extracting Module}
This module processes foregrounds in two steps, i.e., enhancing the real foregrounds via stream merging and removing the counterfeit ones via two filters.

\textbf{Stream Merging}.  Our preliminary experiments suggest that motion of different body parts has varying degrees of impact on different antennas. Therefore, we should merge the CSI data in all the antennas to enhance the impact of motions following a simple rule: one pixel marked as a foreground on any antenna remains so in the merged frame. Otherwise, it is a background.

Stream merging enhances the real foregrounds but also brings more counterfeit ones, which are likely caused by two reasons: device interference and human interference. The former is usually caused by system glitches or other wireless devices competing the same channel, while the latter is caused by nearby human beings.


\textbf{Frequency-temporal Filtering}. This filter deals with the device interference. It leverages a basic fact: human motions usually lasts for a period of time, leading to the temporal correlations of foregrounds. Also, human motion changing the signal propagation pathes usually affects multiple subcarriers, resulting in the frequency correlations of foregrounds. However, a device interference is brief, random and scattered. Therefore, any foreground segment at time $t$ meeting either of the follow criteria will be marked as background,

\begin{itemize}
  \item It lasts less that $\tau$ seconds.
  \item It affects less that $p$ of all subcarriers.
\end{itemize}

Where $\tau$ and $p$ are set to 0.1 and 70\% in SleepGuardian, respectively.

\textbf{ Motion Density Filtering}. This filter handles the human interference. It is built on a general investigation: a non-line-of-sight (NLOS) human motion usually leaves a much lighter impact on channel response than its LOS kind. Therefore, we use a sliding window of 0.5 s length and evaluate the density of foregrounds within the window. If its density is lower than $d$, all foregrounds within the current window will be marked as background. $d$ is set to 0.4 in SleepGuardian.


\subsection{Sleep Guardian Module}

\textbf{Offline Logging}. This services estimates the duration of different sleep postures and the start time, duration and intensity of motions, based on which a fine-grained sleep log is generated including timing and regularity of bed time, onset of sleep, night time awakenings, time of waking up in the mornings, day time naps, and day time sleepiness. It even recognizes which part of the body moves via machine learning.

\textbf{ Online Guarding}. This service keeps guarding the subject for any abnormal behaviors like intensive body twitches from a nightmare, a sudden seizure attack or falling off the bed. Once an abnormality happens, it will automatically warns the designated contacts like a nearby emergency room or a closeby relative.

The abnormality is application-dependent. For instance, an abnormality of an infant could be an unusually high frequency of motions, implying that s/he is awake feeling uncomfortable on bed and the parents should be warned.  For a person having a history of seizure, the abnormality could be a series of periodic motions lasting for a few minutes.

\section{Implementation and Evaluation}

\subsection{Implementation}
\label{subsec:imp}
\begin{figure}[h]
    \centering
   \includegraphics[width=0.8\columnwidth]{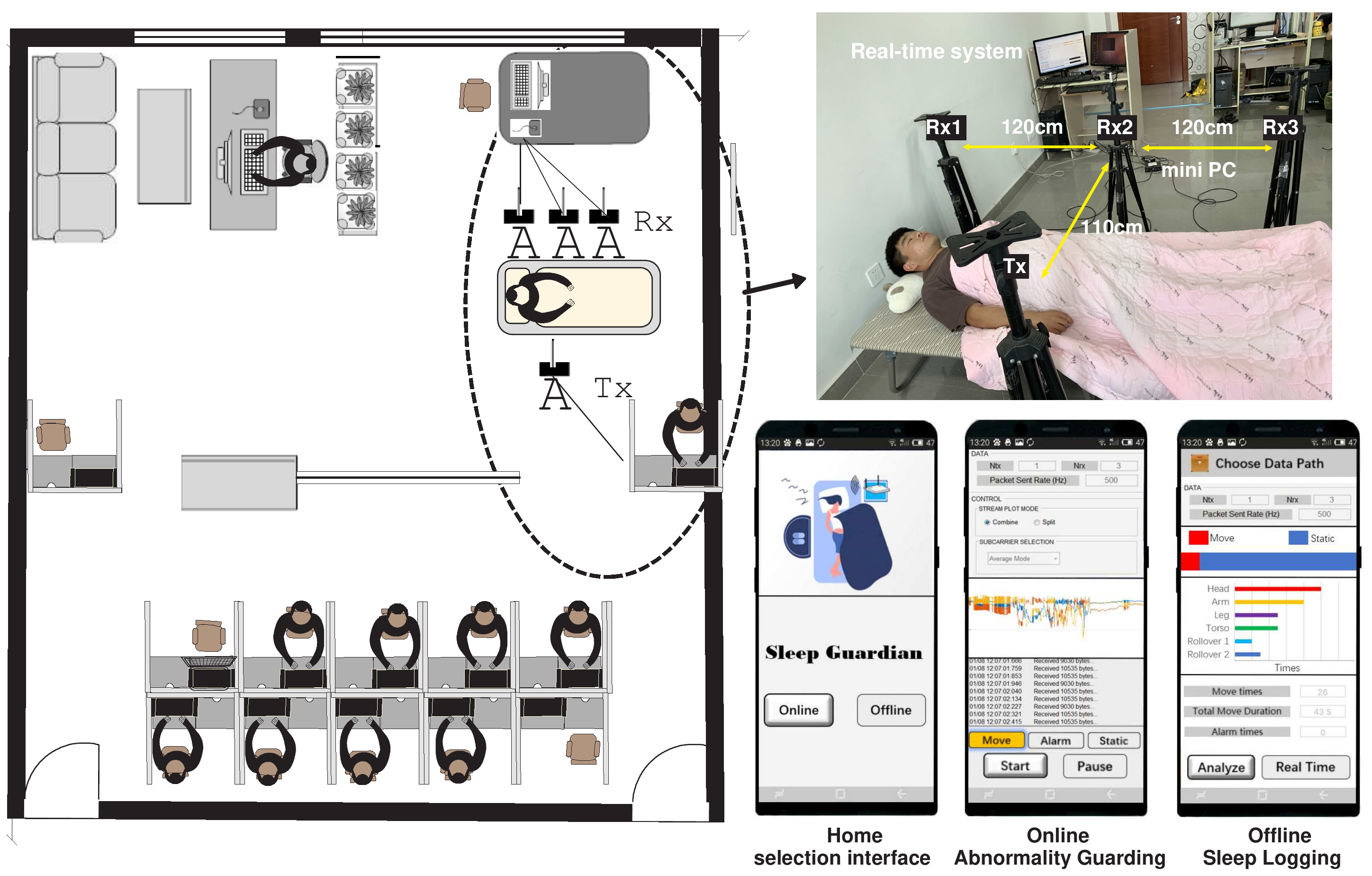}
    \caption{Our prototype SleepGuardian system in a $10\times7 m^2$ office.}
   \label{fig:Layout}
\end{figure}

{\textbf{Prototype System:}} The prototype consists of two low-cost commodity MiniPCs (2.16GH CPU, 4GB RAM and 240GB SSD). One MiniPC is linked with one external antenna as the sender, sending 330 packages per second. The other MiniPC is equipped with three antennas as the receiver. They are both mounted with Intel Network Interface Controller 5300 (5GHz).  Fig. \ref{fig:Layout} shows such experimental setting. The distance between Tx and Rx$2$ is 110cm, while Rx$1$ and Rx$3$ are located left/right side of Rx$2$ with distance 120cm, respectively.

{\textbf{Sleep Guardian Application:}} Sleep is featured by inactivity, and thus SleepGuadian focuses on the motions, especially the abnormal ones, happened during sleep. To understand the key difference between the normal and abnormal motions, Table 1 lists most commonly-seen normal motions that could happen during sleep according to the involving body parts.

We developed a mobile application for users other than the monitored subject. Its interface is shown in Fig.\ref{fig:Layout}. We provide two functions in the application: online abnormality guarding and offline sleep logging. The former focuses on the real-time motion detection, while the latter records the number of motions during sleeping time.

\textbf{\emph{1) Online Abnormality Guarding:}} To ensure a prompt response to any emergency, we leverage edge computing deployed at the local server to detect any abnormal behaviors like intensive body twitches from a nightmare, a sudden seizure attack or falling off the bed. Once an abnormality happens, the server will issue an emergency warning to designated contacts via the mobile application. The interface of online abnormality guarding is shown in Fig.\ref{fig:Layout}.

\textbf{\emph{2) Offline Sleep Logging:}} To maintain a long-term sleep log, we explore machine learning to decode fine-grained movement information as well as the still posture (the corresponding interface is shown in Fig.\ref{fig:Layout}). This function estimates the frequency and duration of each body movement, and records the alert times.


\subsection{Evaluation Setting}
\label{subsec:performance}

To simulate real sleep, we have recruited 15 participants (5 females), whose age, weight, and height range from 20 to 27, 38kg to 95kg, and 1.60m to 1.82m, respectively.  We have selected two typical environments, i.e., a $10\times7$ $m^{2}$ office room and a $4 \times 8$ $m^{2}$  conference room, as shown in Fig. \ref{fig:nlos}. Each participant is asked to perform a specific sequence of $6$ motions $10$ times involving different parts of the body like head, arm, leg and torso while lying on the bed. A camera is used to record the scene simultaneously to serve as the groundtruth. Therefore, the dataset contains $15 \textrm{ (participants)} \times 6 \textrm{ (motions)} \times 10 \textrm{ (cases)} \times 2 \textrm{ (environments)}=1800$ entries.

\textbf{\emph{Evaluation Metrics:}}  To quantify the performance of SleepGuardian, we focus on (1) Detection Rate (DR): the fraction of motions where SleepGuardian correctly detects among all the detected motions, (2) Recognition Rate (RR): the fraction of motions where SleepGuardian correctly recognizes among all the detected motions, (3) Missing Rate (MR): the fraction of motions where SleepGuardian misses among all motions, (4) Mean Absolute Error (MAE): the mean absolute error between the motion duration that SleepGuardian outputs and the groundtruth.

SleepGuardian first leverages GMM to detect a motion and then uses machine learning to recognize it. Therefore, DR and MAE are responsible for evaluating how precisely a motion can be detected by our system. MR is an important metric for online guarding since an undetected motion may lead to a severe consequence. RR is to evaluate how accurately a detected motion can be recognized by our system. It is key to offline logging.

\textbf{\emph{Motion Classifiers:}} We tried several machine learning algorithms including k-NN, SVM and discriminant analysis for motion recognition, and selected k-NN due to its better performance.

\begin{figure}
    \centering
   \includegraphics[width=\columnwidth]{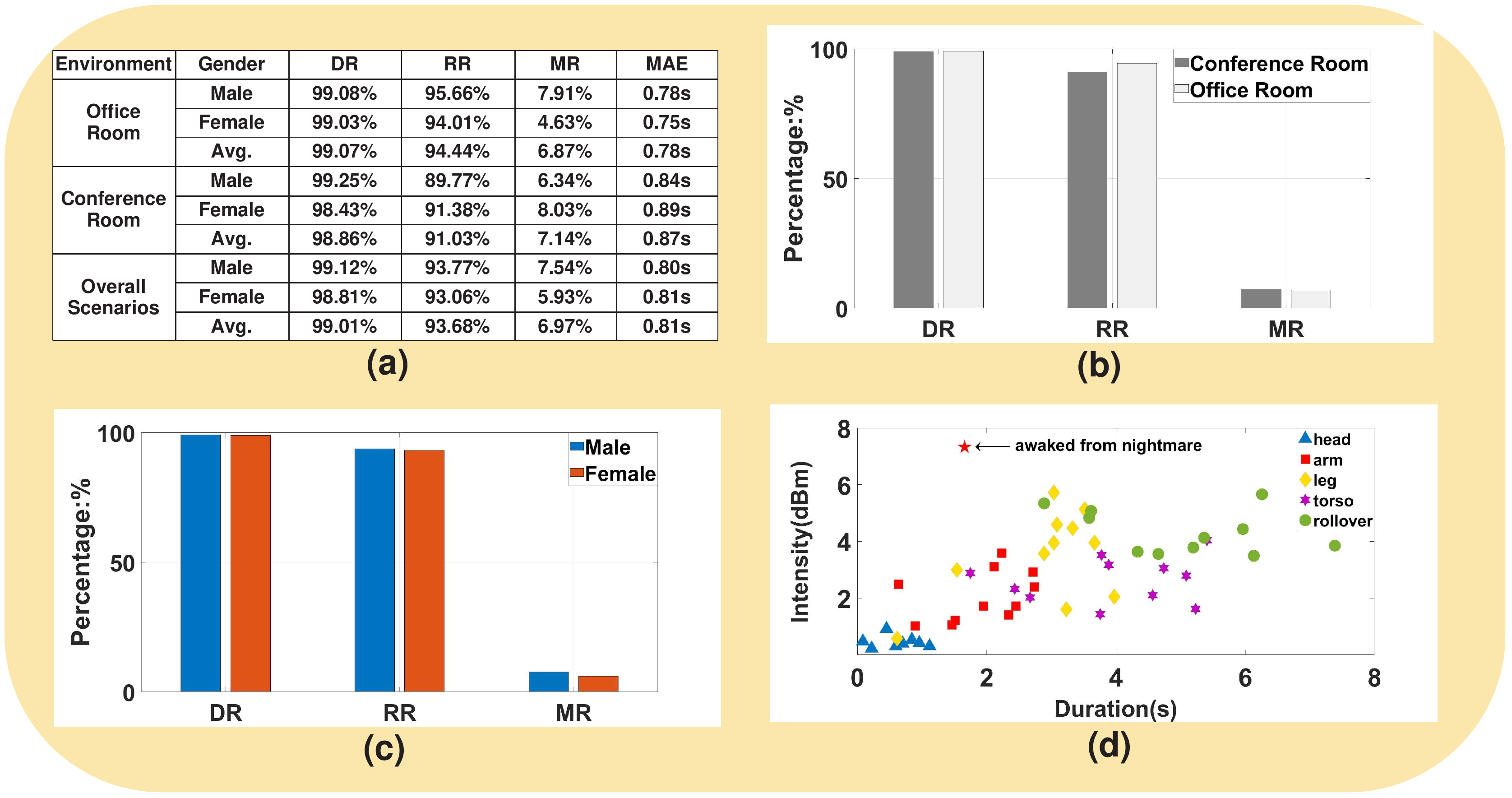}
    \caption{System Evaluation.}
   \label{fig:overallexp}
\end{figure}

\subsection{Performance  Evaluation}

\textbf{\emph{Overall Evaluation}}: Fig. 4(a) presents the statistic results on the real-world experiment: SleepGuardian shows a satisfying performance by achieving $99.01\%$ DR, $93.68\%$ RR, $6.97\%$ MR and $0.81$s MAE on average.

In the following parts, we will interpret the results in terms of environments,  genders, abnormal behavior and body parts.

\textbf{\emph{Environments}}: The office room is not only much larger than the conference room ($70 m^2$ vs $32 m^2$), but also much crowded and noisy with more people. However, such environmental differences make little impact on SleepGuardian, as shown in Fig. 4(b). SleepGuardian achieves an average of $99.07\%$ and $98.86\%$ DR,  $94.44\%$ and $91.03\%$ RR,  $6.87\%$ and $7.14\%$ MR, and  $0.78$s and $0.87$s MAE in the office room and the conference room, respectively. The performance difference is marginal for both environments.

\emph{\textbf{Genders:}} Genders usually lead to differences in the body shape. For instance, the difference in weight of different genders can reach over twice in great in our experiments. However, SleepGuardian shows little performance variation between genders: it achieves an average of $99.12\%$ and $98.81\%$ DR,  $93.77\%$ and $93.06\%$ RR,  $7.54\%$ and $5.93\%$ MR, and  $0.80$s and $0.81$s MAE for male and female, respectively.

{\emph{\textbf{Abnormal behavior:}}} The Sleep Guardian also effectively detects abnormal behaviors by analyzing the duration and intensity (defined as the absolute amplitude change per second) of each motion.  Fig. 4(d) shows one real-world experiment, where a sudden sitting up from a nightmare is detected (red star in Fig. 4(d). The sudden sitting up is isolated from normal motions, and thus easily identified by intensity. We can also observe that motions from the same body part tend to converge, e.g., head movements that are marked by blue triangles, gather together in the $2\times 2$ zone of the graph, owing to the short duration and small intensity during the normal sleeping time.
\begin{table}[]
\caption{The confusion matrix of motion recognition via k-NN}
\begin{tabular}{|c|c|c|c|c|c|c|c|c|}
\hline
Body Part            & Motion      & Head                   & Arm                     & Leg                      & Torso              & Multiple 1         & Multiple 2         & Physical Meaning                        \\ \hline
Head                 & Swing       & 100.0\%                & 0                       & 0                        & 0                  & 0                  & 0                  & Adjusting head position                 \\ \hline
\multirow{2}{*}{Arm} & Up and down & \multirow{2}{*}{6.3\%} & \multirow{2}{*}{93.7\%} & \multirow{2}{*}{0}       & \multirow{2}{*}{0} & \multirow{2}{*}{0} & \multirow{2}{*}{0} & Adjusting the quilt                     \\ \cline{2-2} \cline{9-9}
                     & Swing       &                        &                         &                          &                    &                    &                    & Adjusting arm position                  \\ \hline
\multirow{2}{*}{Leg} & Bend        & \multirow{2}{*}{0}     & \multirow{2}{*}{0}      & \multirow{2}{*}{100.0\%} & \multirow{2}{*}{0} & \multirow{2}{*}{0} & \multirow{2}{*}{0} & \multirow{2}{*}{Adjusting leg position} \\ \cline{2-2}
                     & Stretch     &                        &                         &                          &                    &                    &                    &                                         \\ \hline
Torso                & Twist       & 0                      & 0                       & 12.5\%                   & 87.5\%             & 0                  & 0                  & Lightening the tension                  \\ \hline
Multiple 1           & Rollover    & 0                      & 0                       & 12.5\%                   & 6.3\%              & 81.2\%             & 0                  & Changing postures                       \\ \hline
multiple 2           & Stretch     & 0                      & 0                       & 0                        & 0                  & 0                  & 100.0\%            & Adjusting posture                       \\ \hline
Avg.                 & \multicolumn{6}{c|}{}                                                                                                               & 93.68\%            &                                         \\ \hline
\end{tabular}
\label{table:motion}
\end{table}


\emph{\textbf{Body parts:}} Table \ref{table:motion} illustrates the performance of motion classification. We can observe that SleepGuardian achieves an average accuracy up to 93.68\% using the k-NN algorithm. Most of the normal motions can be identified with high accuracy, however, the rollover sometimes is misidentified as torso. Movement with multiple motions like rollover sometimes is misidentified as the movement of torso, because the intensity and duration of the torso movement are very similar to multiple movements.

In summary, the experimental results demosntrate that SleepGuardian is resilient to the environments, genders and body parts, and thus constitutes a promising solution for sleep monitoring.

\begin{figure}
    \centering
   \includegraphics[width=\columnwidth]{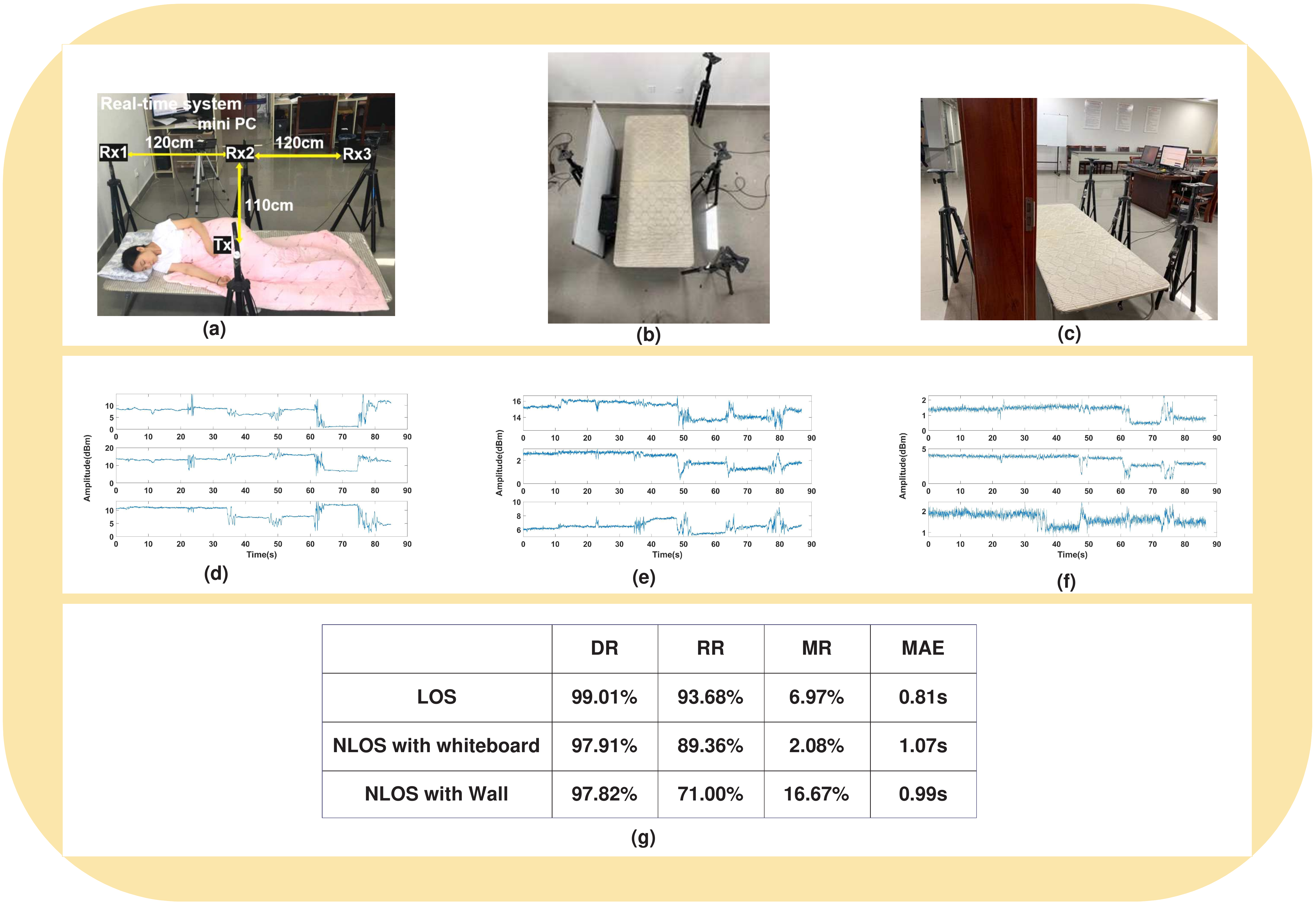}
    \caption{Comparison of the performance in LOS and NLOS scenarios.}
   \label{fig:nlos}
\end{figure}

\subsection{Further Discussions}

 WiFi signal performs differently in LOS and NLOS scenarios. To study SleepGuardian's performance in NLOS scenarios, we perform extra experiments and the results are shown in Fig. \ref{fig:nlos}. The experimental setup is the same as introduced in Section \ref{subsec:imp} and \ref{subsec:performance}, with the only difference of adding a white board and a wall between the sender and the receiver to simulate the NLOS phenomenon. Fig. \ref{fig:nlos} (a)-(c) list the evaluation environment and (d)-(f) show the amplitude in each scenario. The overall detection performance is shown in Fig. \ref{fig:nlos} (g).

From the results, we can observe that the NLOS does affect the overall performance in terms of the DR, RR, MR and MAE, and the wall (which blocks the signal more) affects more than the whiteboard. Please note that the DR and RR in all the three cases are all above 89$\%$, indicating the SleepGuardian's ability to achieve satisfactory performance in the NLOS scenarios.

We are also interested in the theory behind, especially the optimal setup for the NLOS scenarios, which requires further efforts to explore and is left as future work.


\section{Conclusion}
\label{Sect:Con&FutWork}
In this paper, we present SleepGuardian, a RF-based sleep monitoring system leveraging the off-the-self WiFi infrastructure. Unlike its rivals relying on specialized hardware or complicated training, it explores the channel characteristics and models the energy distribution over subcarriers as a Gaussian Mixture Model, which requires no calibrations or target-dependent training to detect normal or abnormal motions during sleep. SleepGuardian has been extensively evaluated in real environments from different perspectives and the results confirmed its reliability and efficiency. SleepGuardian can be seamlessly integrated in existing WiFi infrastructures, and thus constitutes a practical solution in real world.

\section*{Acknowledgments}
The authors would like to thank the anonymous reviewers for their time and effort. This work is sponsored by the National Natural Science Foundation of China (NSFC) under Grant No. 61772169, National Key Research and Development Program under Grant No.2018YFB0803403, the Fundamental Research Funds for the Central Universities under No.JZ2018HGPA0272, and  Open Projects by Jiangsu Province Key Laboratory of Internet of Things under No.JSWLW-2017-002.

\bibliographystyle{IEEEtran}
\bibliography{LoMD}

\begin{thebibliography}{10}
\providecommand{\url}[1]{#1}
\csname url@samestyle\endcsname
\providecommand{\newblock}{\relax}
\providecommand{\bibinfo}[2]{#2}
\providecommand{\BIBentrySTDinterwordspacing}{\spaceskip=0pt\relax}
\providecommand{\BIBentryALTinterwordstretchfactor}{4}
\providecommand{\BIBentryALTinterwordspacing}{\spaceskip=\fontdimen2\font plus
\BIBentryALTinterwordstretchfactor\fontdimen3\font minus
  \fontdimen4\font\relax}
\providecommand{\BIBforeignlanguage}[2]{{%
\expandafter\ifx\csname l@#1\endcsname\relax
\typeout{** WARNING: IEEEtran.bst: No hyphenation pattern has been}%
\typeout{** loaded for the language `#1'. Using the pattern for}%
\typeout{** the default language instead.}%
\else
\language=\csname l@#1\endcsname
\fi
#2}}
\providecommand{\BIBdecl}{\relax}
\BIBdecl

\bibitem{BLS2016}
U.~B. of~Labor~Statistics, ``Charts from the american time use survey,''
  Website, 2016, \url{https://www.bls.gov/tus/charts/}.

\bibitem{Seelye2015The}
A.~Seelye, N.~Mattek, D.~Howieson, T.~Riley, K.~Wild, and J.~Kaye, ``The impact
  of sleep on neuropsychological performance in cognitively intact older adults
  using a novel in-home sensor-based sleep assessment approach.''
  \emph{Clinical Neuropsychologist}, vol.~29, no.~1, pp. 53--66, 2015.

\bibitem{Sadeh1995The}
A.~Sadeh, P.~J. Hauri, D.~F. Kripke, and P.~Lavie, ``The role of actigraphy in
  the evaluation of sleep disorders.'' \emph{Sleep}, vol.~18, no.~4, pp.
  288--302, 1995.

\bibitem{Barsocchi2017An}
P.~Barsocchi, M.~Bianchini, A.~Crivello, D.~L. Rosa, F.~Palumbo, and
  F.~Scarselli, ``An unobtrusive sleep monitoring system for the human sleep
  behaviour understanding,'' in \emph{IEEE International Conference on
  Cognitive Infocommunications}, Oct 2017.

\bibitem{Da2014Nocturnal}
W.~J. Da, H.~H. Su, H.~N. Yoon, Y.~J. Lee, D.~U. Jeong, and K.~S. Park,
  ``Nocturnal awakening and sleep efficiency estimation using unobtrusively
  measured ballistocardiogram.'' \emph{IEEE transactions on bio-medical
  engineering}, vol.~61, no.~1, pp. 131--8, 2014.

\bibitem{Abbas2011Neonatal}
A.~K. Abbas, K.~Heimann, K.~Jergus, T.~Orlikowsky, and S.~Leonhardt, ``Neonatal
  non-contact respiratory monitoring based on real-time infrared
  thermography,'' \emph{BioMedical Engineering OnLine}, vol.~10, no.~1, p.~93,
  2011.

\bibitem{Wang2013Unconstrained}
C.~W. Wang, A.~Hunter, N.~Gravill, and S.~Matusiewicz, ``Unconstrained video
  monitoring of breathing behavior and application to diagnosis of sleep
  apnea.'' \emph{IEEE transactions on bio-medical engineering}, vol.~61, no.~2,
  pp. 396--404, 2013.

\bibitem{Li2016Non}
M.~H. Li, A.~Yadollahi, and B.~Taati, ``Non-contact vision-based
  cardiopulmonary monitoring in different sleeping positions,'' \emph{IEEE
  Journal of Biomedical \& Health Informatics}, pp. 1--1, 2016.

\bibitem{sigg2014telepathic}
S.~Sigg, U.~Blanke, and G.~Troster, ``The telepathic phone: Frictionless
  activity recognition from wifi-rssi,'' in \emph{Proc. of the IEEE PERCOM
  2014}, Budapest, Hungary, March 2014, pp. 148--155.

\bibitem{Patwari2011Monitoring}
N.~Patwari, J.~Wilson, S.~Ananthanarayanan, S.~K. Kasera, and D.~R. Westenskow,
  ``Monitoring breathing via signal strength in wireless networks,'' \emph{IEEE
  Transactions on Mobile Computing}, vol. abs/1109.3898, no.~8, pp. 1--1, 2011.

\bibitem{Liu2014WiSleep}
X.~Liu, J.~Cao, S.~Tang, and J.~Wen, ``Wi-sleep: Contactless sleep monitoring
  via wifi signals,'' in \emph{Proc. of IEEE Real-Time Systems Symposium
  (RTSS)}, Rome, Italy, Dec 2014, pp. 346--355.

\bibitem{Liu2016Contactless}
X.~Liu, J.~Cao, S.~Tang, J.~Wen, and P.~Guo, ``Contactless respiration
  monitoring via off-the-shelf wifi devices,'' \emph{IEEE Transactions on
  Mobile Computing}, vol.~15, no.~10, pp. 2466--2479, Oct 2016.

\bibitem{Gu2018Sleepy}
Y.~Gu, Y.~Zhang, J.~Li, Y.~Ji, X.~An, and F.~Ren, ``Sleepy: Wireless channel
  data driven sleep monitoring via commodity wifi devices,'' \emph{IEEE
  Transactions on Big Data}, 10.1109/TBDATA.2018.2851201, 2018.

\bibitem{Hossain2018EnClass}
M.~S. Hossain and G.~Muhammad, ``Environment classification for urban big data
  using deep learning,'' \emph{IEEE Communications Magazine}, vol.~56, no.~11,
  pp. 44--50, November 2018.

\bibitem{Zheng2016Smokey}
X.~Zheng, J.~Wang, L.~Shangguan, Z.~Zhou, and Y.~Liu, ``Smokey: Ubiquitous
  smoking detection with commercial wifi infrastructures,'' in \emph{Proc. of
  IEEE INFOCOM 2016}, Hong Kong, April 2015, pp. 17--18.

\end{thebibliography}

\begin{IEEEbiography}[{\includegraphics[width=1in,height=1.25in,clip,keepaspectratio]{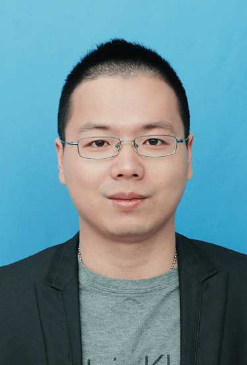}}]{Yu Gu} (M'10-SM'16)received the B.E. degree from the Special Classes for the Gifted Young, University of Science and Technology of China, Hefei, China, in 2004, and the D.E. degree from the same university in 2010.In 2006, he was an Intern with Microsoft Research Asia, Beijing, China, for seven months. From 2007 to 2008, he was a Visiting Scholar with the University of Tsukuba, Tsukuba, Japan. From 2010 to 2012, he was a JSPS Research Fellow with the National Institute of Informatics, Tokyo, Japan. He is currently a Professor and Dean Assistant with the School of Computer and Information, Hefei University of Technology, Hefei, China. His current research interests include pervasive computing and affective computing. He was the recipient of the IEEE Scalcom2009 Excellent Paper Award, NLP-KE2017 Best Paper Award and IEEE CCIS2018 Best Student Paper Award. He is a senior member of IEEE and CCF.
\end{IEEEbiography}

\begin{IEEEbiography}[{\includegraphics[width=1in,height=1.25in,clip,keepaspectratio]{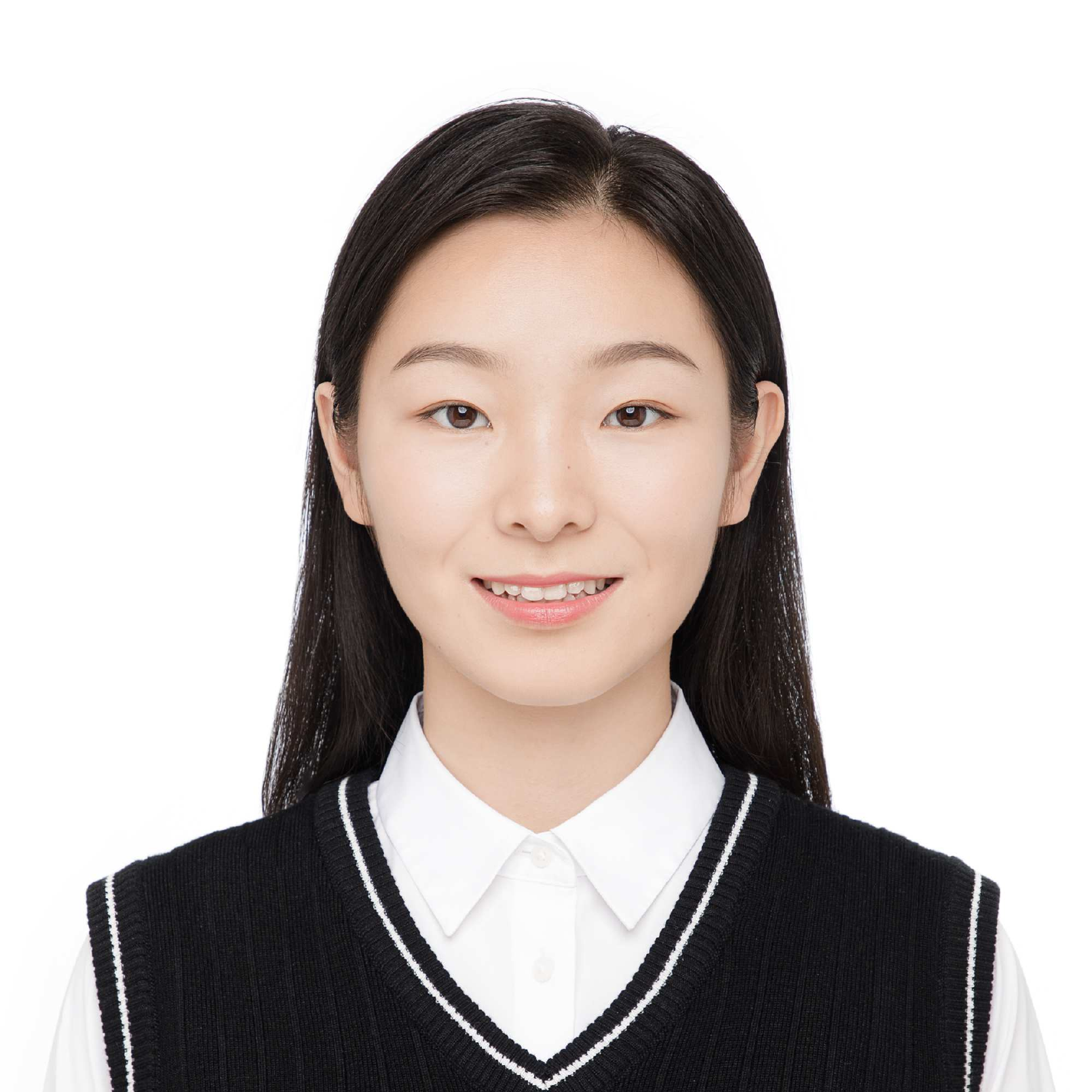}}]{Yantong Wang} ((M'19) received the B.E degree from the Shanghai Normal University in 2016. From 2017 to now, she is a postgraduate student in the Hefei University of Technology. Her research interest includes affective computing and sensorless sensing.
\end{IEEEbiography}

\begin{IEEEbiography}[{\includegraphics[width=1in,height=1.25in,clip,keepaspectratio]{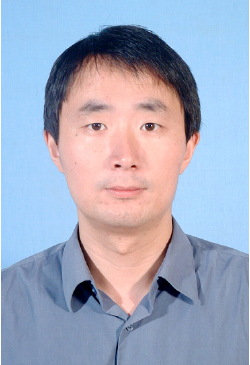}}]{
Jun Liu} received the B.E.degree from North East Normal University, Changchun, China, in 2002, and M.E. degree from Hefei University of Technology, Hefei, China, in 2007, and Ph.D degree from the Institute of Computing Technology, Chinese Academy of Science in 2011. From Nov. 2016 to Dec. 2017, he was a Visiting Scholar with the University of Texas at Austin. He currently is an associate professor with the School of Computer and Information, Hefei University of Technology. His research interests include effective computing, computer architecture. ?
\end{IEEEbiography}

\begin{IEEEbiography}[{\includegraphics[width=1in,height=1.25in,clip,keepaspectratio]{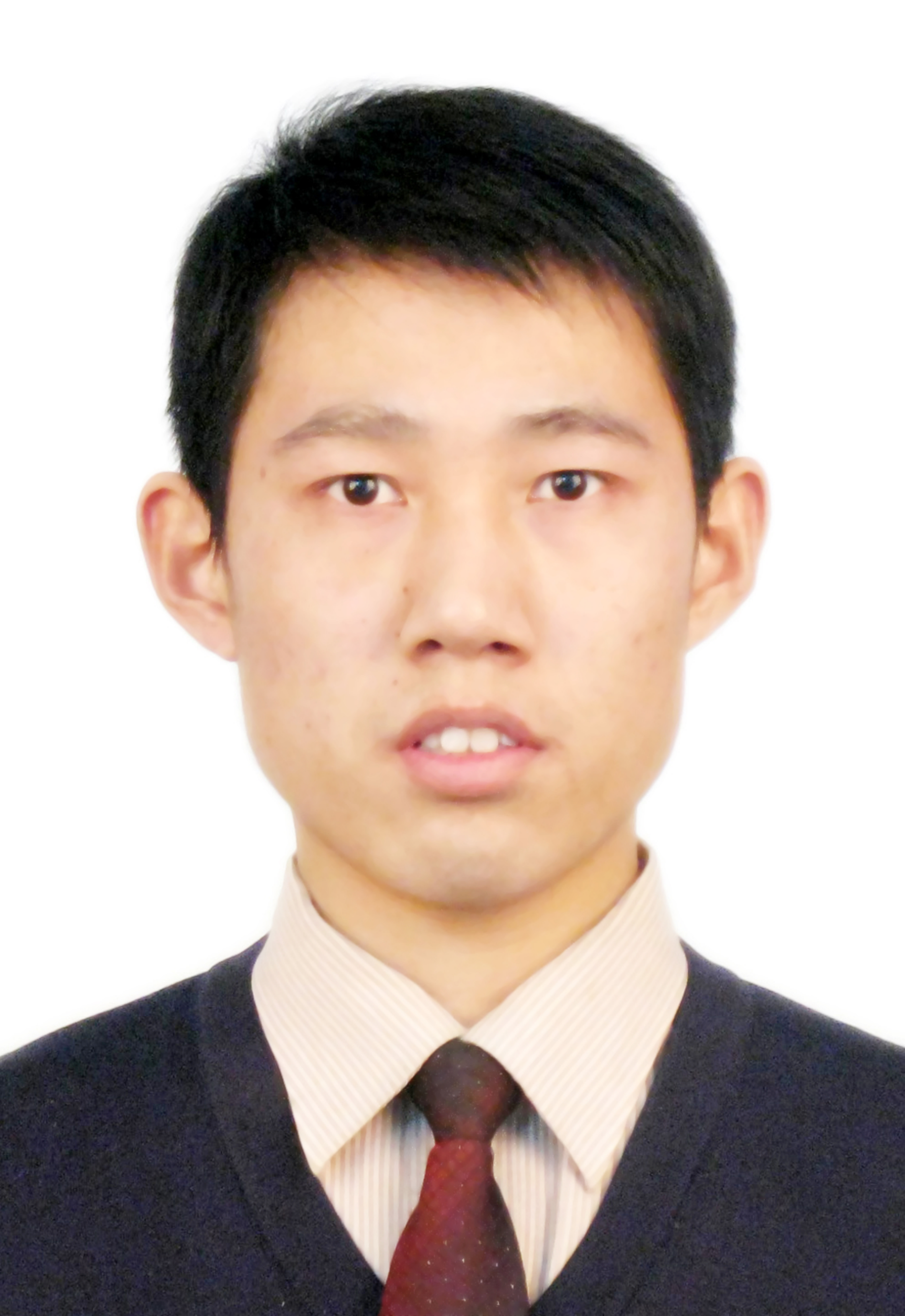}}]{Zhi Liu} received the B.E., in computer science and technology from the University of Science and Technology of China, China, in 2009 and Ph.D. degree in informatics in National Institute of Informatics and The Graduate University for Advanced Studies (Sokendai) Tokyo, Japan. He is currently an Assistant Professor at Shizuoka University and an adjunct researcher at Waseda University, Japan. He was a Junior Researcher (Assistant Professor) at Waseda University from Dec. 2104 to Mar. 2017, and a JSPS research fellow in National Institute of Informatics and The Graduate University for Advanced Studies (Sokendai) from Apr. 2012 to Nov. 2014. From Oct.2009 to Mar. 2014, he was a research assistant in National Institute of Informatics and The Graduate University for Advanced Studies (Sokendai). His research interest includes wireless networks, video/image processing and transmission. He was the recipient of the IEEE StreamComm2011 best student paper award, VTC2014-Spring Young Researchers Encouragement Award, 2015 IEICE Young Researcher Award and ICOIN2018 best paper award. He is a member of IEEE and IEICE.
\end{IEEEbiography}

\begin{IEEEbiography}[{\includegraphics[width=1in,height=1.25in,clip,keepaspectratio]{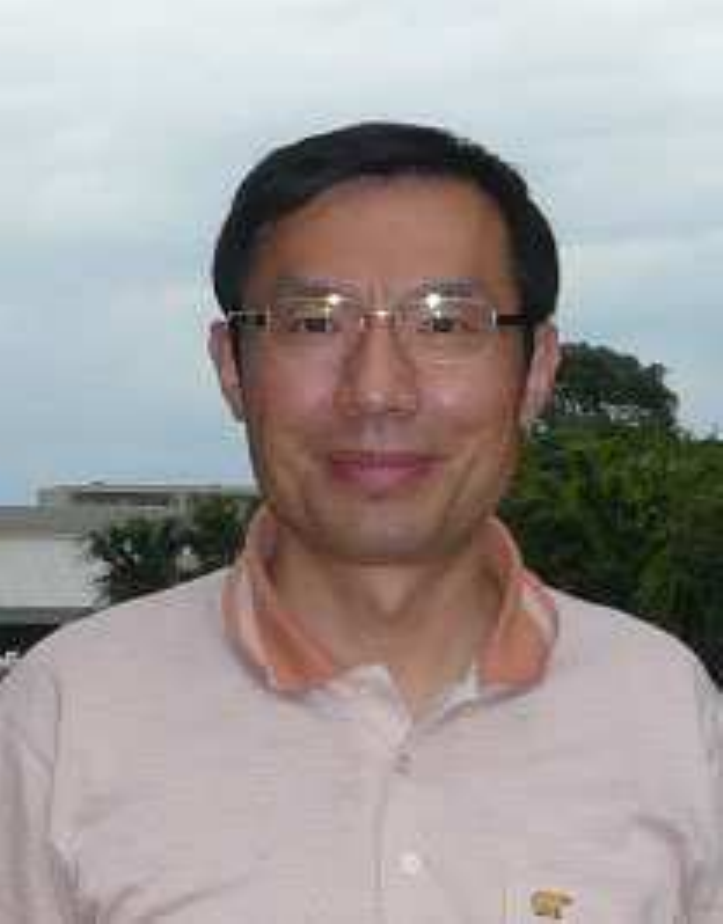}}]{Jie Li} received the B.E. degree in computer science from Zhejiang University, Hangzhou, China, the M.E. degree in electronic engineering and communication systems from China Academy of Posts and Telecommunications, Beijing, China. He received the Dr. Eng. degree from the University of Electro-Communications, Tokyo, Japan. He is with Faculty of Engineering, Information and Systems, University of Tsukuba, Japan, where he is a Professor. His current research interests are in mobile distributed computing and networking, big data and cloud computing, network security, OS, modeling and performance evaluation of information systems. He was a visiting Professor in Yale University, USA, Inria Sophia Antipolis, France and Inria Grenoble - Rhone-Aples, France. He is a senior member of IEEE and ACM and a member of IPSJ (Information Processing Society of Japan). He is the Chair of Technical Committee on Big Data (TCBD), IEEE Communications Society. He has served as a secretary for Study Group on System Evaluation of IPSJ and on several editorial boards for the international Journals, and on Steering Committees of the SIG of System EVAluation (EVA) of IPSJ, the SIG of DataBase System (DBS) of IPSJ, and the SIG of MoBiLe computing and ubiquitous communications of IPSJ. He has also served on the program committees for several international conferences such as IEEE INFOCOM, IEEE GLOBECOM, and IEEE MASS.
\end{IEEEbiography}

\end{document}